\documentclass[prb,aps,twocolumn,groupedaddress,floats,nopacks,final,superscriptaddress]{revtex4-1}
\usepackage{graphicx}
\usepackage{bm}
\usepackage{color}
\usepackage{epsfig}
\usepackage{psfrag}
\usepackage{amsmath,amsfonts,amssymb,bm,ulem}

\begin{document}

\author{Igor S. Tupitsyn}
\affiliation{Department of Physics, University of Massachusetts, Amherst, MA 01003, USA}
\affiliation{Russian Research Center ``Kurchatov Institute'', 123182 Moscow, Russia}
\author{Nikolay V. Prokof'ev}
\affiliation{Department of Physics, University of Massachusetts, Amherst, MA 01003, USA}
\affiliation{Russian Research Center ``Kurchatov Institute'', 123182 Moscow, Russia}


\title{Orthogonality catastrophe in Coulomb systems}



\begin{abstract}
The orthogonality catastrophe (OC) problem is considered solved for fifty years.
It has important consequences for numerous dynamic phenomena in fermionic systems,
including Kondo effect, X-ray spectroscopy, and quantum diffusion of impurities,
and is often used in the context of metals. However, the key assumptions on which
the known solution is based---impurity potentials with finite cross-section and
non-interacting fermions---are both highly inaccurate for problems involving charged particles in metals. As far as we know, the OC problem for the ``all Coulomb" case has never been addressed systematically, leaving it unsolved for the most relevant practical applications. In this work we
include effects of dynamic screening in a consistent way and demonstrate that for short-range impurity potentials the non-interacting Fermi-sea approximation radically overestimates the power-law decay exponent of the overlap integral. We also find that the dynamically screened Coulomb potential
leads to a larger exponent than the often used static Yukawa potential. Finally, by employing the Diagrammatic Monte Carlo technique, we quantify effects of a finite impurity mass and reveal how OC physics leads to small, but finite, impurity residues.
\end{abstract}

\maketitle

\section{Introduction}
A prototypical process leading to the Anderson orthogonality catastrophe (OC) problem is a sudden (at time $t=0$) excitation of a core electron in an atom, as in the X-Ray absorption (XAS, see, for instance, Ref.~\cite{XAS}), leaving a hole in a deep core level. In the so-called ``hard X-ray" limit the electron leaves the sample---this case is very convenient for studying the OC problem \cite{Anderson1967}. The localized (i.e. of infinite mass $M$) core-hole acts in the same way as a heavy impurity in Anderson's formulation: its potential polarizes the surrounding vacuum by creating electron-hole pairs around the Fermi level. The decay of the overlap integral modulus squared,
$I(t)=|\langle \Psi (0) | \Psi (t) \rangle|^2$, between the initial system's state and the state at time $t>0$ is directly related to singular properties of the X-ray spectra near the threshold \cite{Mahan1967,Anderson1967,Nozier-I,Nozier-II,Nozier-III}.

In the standard approach to the OC problem \cite{Nozier-I,Nozier-II,Nozier-III} the impurity potential, $V_S$, is assumed to have a finite scattering cross-section. [In what follows we will keep using the notion of ``impurity" regardless of its physical origin.] Indeed, the exponent controlling the power law decay of $I(t)$ is given by $\gamma =  2\sum_l (2 l + 1) (\delta_l/\pi)^2$, where $\delta_l$ is the scattering phase shift in the orbital channel $\ell$ at the Fermi energy. It is finite if, and only if, the scattering cross-section is finite. The other simplifying assumption is that particles and holes near the Fermi level are non-interacting; it is justified by the quasi-particle picture of the Fermi liquid state emerging at low temperature.

\begin{figure}[tbh]
\includegraphics[width=0.95\columnwidth]{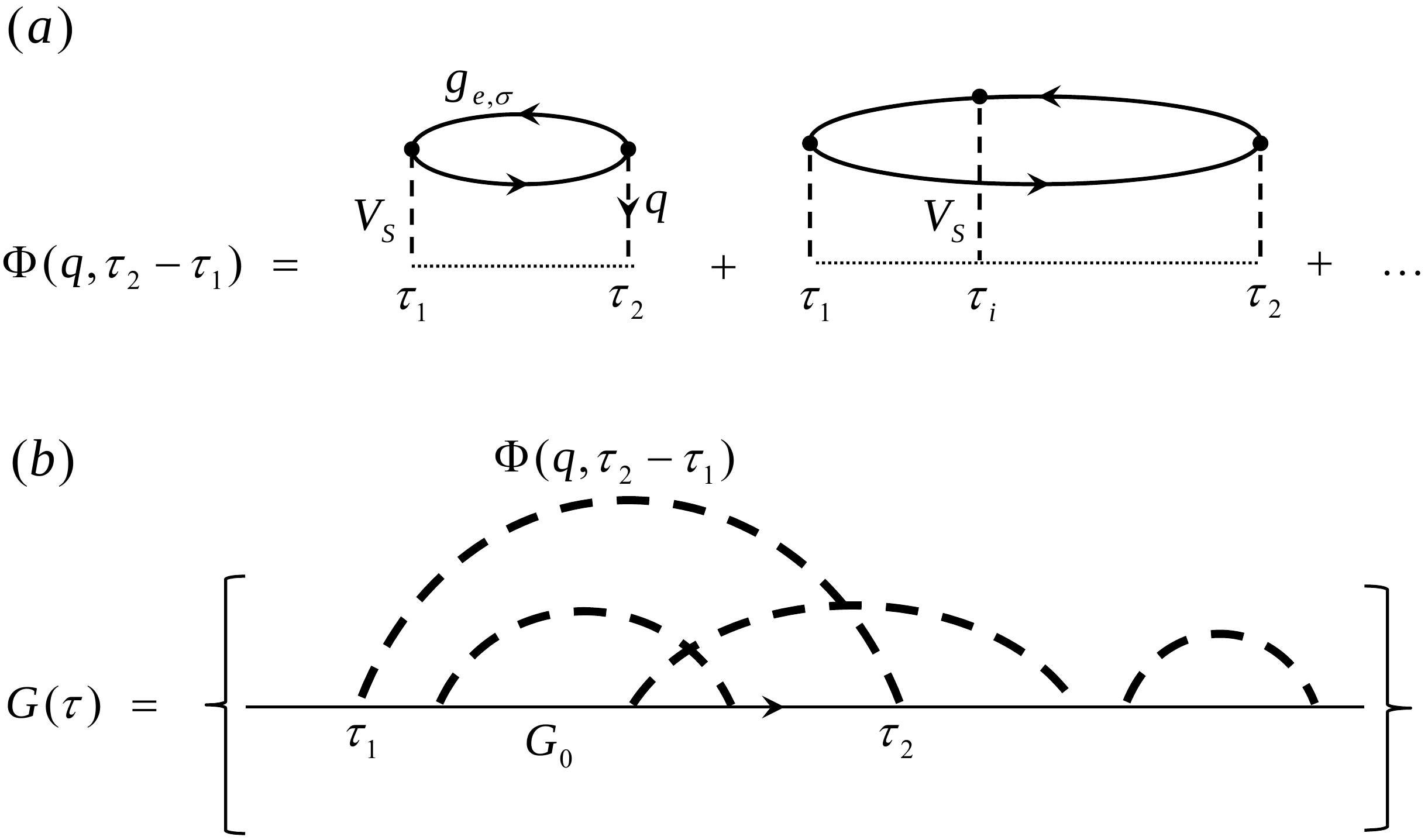}
\caption{(color online). (a) One-loop diagrammatic contributions to the impurity self-energy $\Phi({\bf q},\tau_2-\tau_1)$  in the imaginary time representation. The summation/integration over all possible intermediate scattering events on the time interval $(\tau_1,\tau_2)$ is assumed. $V_S$ is the impurity potential with finite scattering cross-section, and $g_{e,\sigma}$ is the Green's function of electrons with spin $\sigma$ in the Fermi-sea. (b) The impurity Green's function, $G(\tau)$, is obtained by expanding in the number of one-loop contributions and integrating over their parameters. $G_0(\tau)$ is the bare impurity Green's function. }
\label{fig1}
\end{figure}

Scattering of electrons and holes off the impurity potential $V_S$ can be visualized in
terms of Feynman diagrams, see Fig.~\ref{fig1}. For the localized impurity, any diagram
can be decomposed into the product of independent one-loop contributions because for
all intermediate states $G_{0}(\tau) = \exp{ (- E_0 \tau) }$, where $E_0$ is the bare impurity ``energy" ($E_0$ absorbs the Hartree term, not shown in Fig.~\ref{fig1}(a), and in what follows we set $E_0$ to zero). [Note that this decomposition is no longer valid for impurity with finite mass $M$ because the bare Green's functions in intermediates states depend on the momentum transfer to the bath.] One-loop diagrams are based on the Taylor series expansion in powers of $V_S$, and account for arbitrary number of intermediate scattering events on the time interval $(\tau_1, \tau_2)$ for both the electron and hole, see Fig.~\ref{fig1}(a). It is convenient to represent one-loop contributions with an equivalent bosonic propagator $\Phi({\bf q}, \tau)$. Summation over independent one-loop contributions to the impurity Green's function, see Fig.~\ref{fig1}(b), immediately leads to the exponential form (see Ref.~\cite{Nozier-III})
\begin{eqnarray}
&&G(\tau) = \exp \left\{ - \int^{\tau}_0  \int^{\tau_2}_0 d\tau_1 d\tau_2\;
\digamma (\tau_2-\tau_1) \right\}; \;\;\;\;\;\; \label{Gh_Inf_1} \\
&&\digamma(\tau_2-\tau_1) = \int \frac{d{\bf q}}{(2 \pi)^3} \; \Phi({\bf q},\tau_2-\tau_1).
\label{Gh_Inf_2}
\end{eqnarray}

In the long-time limit $\tau >> 1/\epsilon_F$, where $\epsilon_F$ is the Fermi energy, an exact solution for $\digamma$ function in the standard approach \cite{Nozier-III} features a power law decay, $\digamma(\tau ) \to - \gamma \tau^{-2}$. This result immediately implies that if we express the impurity Green's function at long times in terms of impurity energy, $E$, and $Z$-factor as
\begin{equation}
G(\tau)= Z(\tau)e^{-E \tau}, \qquad (\tau \to \infty)\,,
\label{G_Z}
\end{equation}
then $Z(\tau ) = I(\tau) \propto \tau^{- \gamma}$.

Assumptions on which the standard approach is based are well for experimental setups with weakly interacting ultra-cold fermions \cite{Demler-OC-CA,Demler-OC-HI}.
However, they are invalid for XAS (as well as for Resonant Inelastic/Soft X-ray Spectroscopy \cite{RIXS-REW-1,RIXS-REW-2,Demler-RSXS}) in metals or, more generally, for any problem involving charged Fermi liquids and impurities. To begin with, for charged impurities the Coulomb potential, $V_C({\bf q})=4 \pi e^2 / q^2$, has infinite cross-section, and if one were to formally replace $V_S$ with $V_C$ in Fig.~\ref{fig1}, the $\digamma$ function would feature a divergent integral over momentum transfer---this would literally constitute an orthogonality ``disaster" invalidating the solution in terms of the Fermi-surface phase shifts \cite{Nozier-I,Nozier-II,Nozier-III}. Thus, one cannot avoid considering electrons in the Fermi-sea as interacting via the Coulomb potential because otherwise the impurity potential cannot be screened. Finally, even for short-range impurity potentials effects of dynamic screening in metals remain non-perturbative (high-order bubble-type diagrams diverge, and the entire geometrical series needs to be summed up), and any treatment ignoring them is highly inaccurate. 

\begin{figure}[tbh]
\includegraphics[width=0.95\columnwidth]{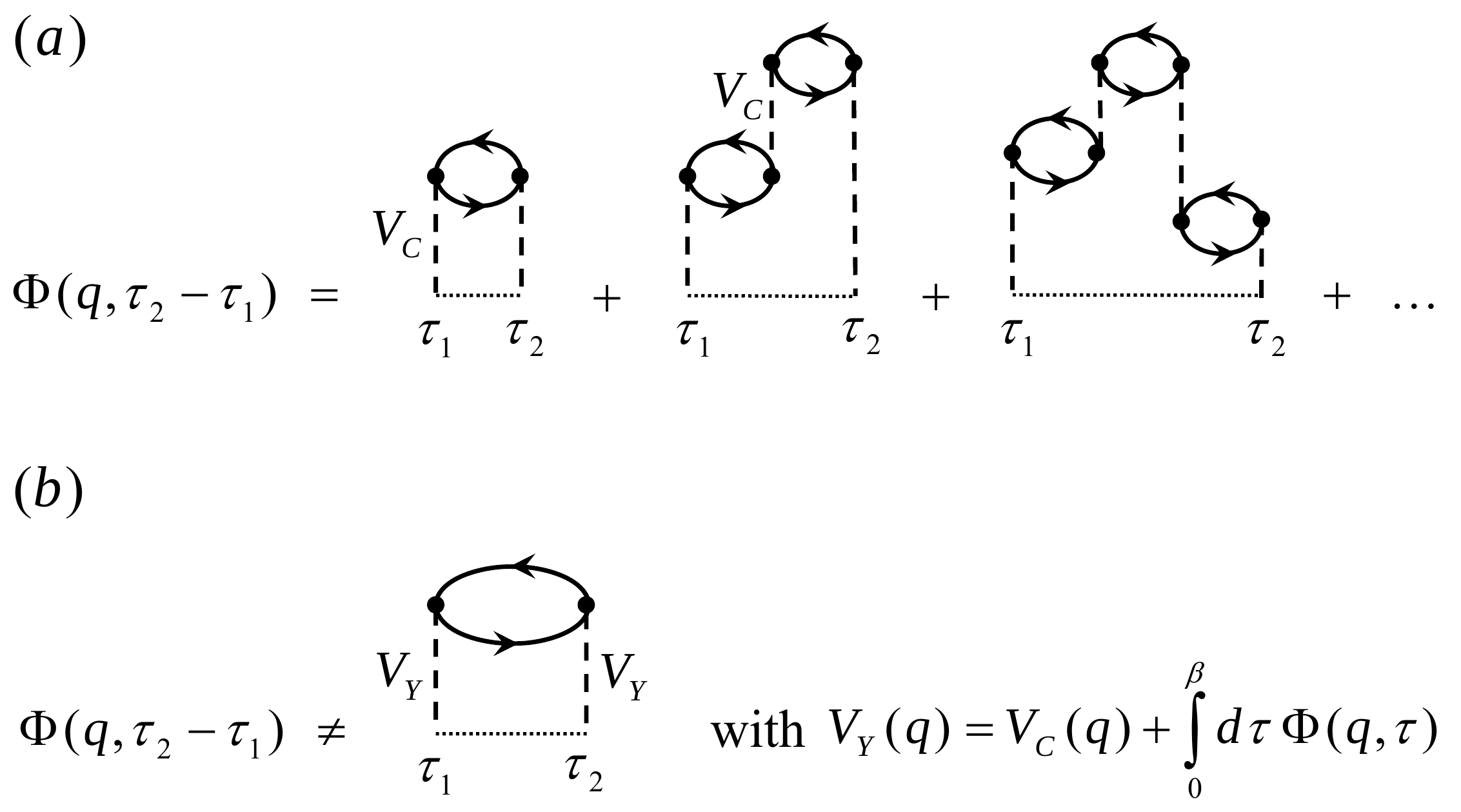}
\caption{(color online). (a) Diagrams leading to dynamic screening of the Coulomb potential.
For perturbative values of $r_s$ they provide the dominant contribution to the $\Phi$ function. (b) Substituting Coulomb potential with the static screened potential, $V_Y$,
leads to inconsistent treatment and multiple counting of bubble insertions.}
\label{fig2}
\end{figure}

As we illustrate in Fig.~\ref{fig2}, replacing the Coulomb potential with the {\it static screened potential}, $V_Y({\bf q})=V_C({\bf q})/\varepsilon({\bf q},\omega=0)$, where $\varepsilon$ is the dielectric function, or a simpler Yukawa form, $V_Y({\bf q})=4\pi e^2/(q^2+\kappa^2)$, where $\kappa$ is the Thomas-Fermi momentum (see, for instance, \cite{Nozier-I,Nozier1971}), is mathematically inconsistent and leads to multiple counting of bubble insertions. Indeed, static properties may not appear in the dynamic formulation of the problem where the system is allowed to evolve only for a finite amount of time. Moreover, the $\Phi$ function in Fig.~\ref{fig2}(a) is based on a single geometric series; an attempt to replace it with the diagram shown in Fig.~\ref{fig2}(b) introduces two geometric series of identically the same nature connected by an element on which these series are built. Therefore, to properly account for {\it dynamic screening} effects one has to deal with the $\Phi$ function
\begin{equation}
\Phi({\bf q},\omega) = V_C({\bf q}) (\varepsilon^{-1}({\bf q},\omega) - 1) \;,
\label{Gh_V}
\end{equation}
based on the dynamic dielectric response $\varepsilon$.

It is clear that dynamic screening will remove the spurious divergence of the momentum integral in $\digamma$. Since OC originates from response of gapless particle-hole excitations near the Fermi-surface, and these are present in the metallic Fermi-liquid state, it is also expected that the power-law OC scenario does take place \cite{Nozier-I,Nozier1971}). However, to which extent the collective plasmon excitations, also existing in Coulomb systems, modify the OC exponent is far from obvious. 

In this work, we first consider the response of plasmon modes alone, and show that within the plasmon-pole approximation (PPA) to $\varepsilon$ (for original formulation see, for instance, Refs.~\cite{Lundquist67-1,Lundquist67-2,HedLund1969}), the OC is eliminated and the impurity $Z$ factor saturates to a constant in the limit of $\tau \to \infty$. Next, we address the OC problem within the random-phase approximation (RPA), see Fig.~\ref{fig2}(a), which becomes exact in the limit of small Coulomb parameter $r_s$ and, correspondingly, takes into account both gapped and gapless Fermi-liquid modes. In RPA, the OC in its canonical power-law form is restored, but screening effects dramatically alter the value of the exponent $\gamma$. Even for short-range impurity potentials, the non-interacting Fermi-sea approximation fails to produce reasonable results for metals.
By comparing the power-law decay obtained for the dynamically screened Coulomb potential
with that for the often used, but formally inconsistent, scheme combining the static Yukawa impurity potential with the non-interacting Fermi-sea approximation, we find that the latter is characterized by a smaller exponent. Finally, we employ the diagrammatic Monte Carlo (DiagMC) technique for polarons, introduced earlier in Refs.~\cite{Polaron1998,Polaron2000}, to compute the Green's function of mobile impurities (i.e., with finite mass) and reveal how the OC is truncated by recoil effects.

\section{Formalism}

To calculate the impurity Green's function, $G(\tau)$, and obtain its $Z$-factor in the limit of infinite mass $M$, we use expressions (\ref{Gh_Inf_1})-(\ref{Gh_V}), which provide an exact solution to the problem in the limit of small $r_s$. For the $\Phi$ function, Eq.~(\ref{Gh_V}), unless stated differently, we either use the PPA, or the RPA expressions, derived for the jellium model. Since the $\Phi$ function is based on the geometric series, it can be obtained numerically very efficiently with the use of fast Fourier transforms.

For finite $M$ we employ the DiagMC technique for polarons in the Matsubara momentum - imaginary time representation. This technique allows one to obtain the impurity Green's function by unbiased sampling of the configuration space of Feynman's diagrams illustrated in Fig.~\ref{fig1}(b) (for details see Refs.~\cite{Polaron1998,Polaron2000}). More specifically, for bare impurity and free electron Green's functions we consider
\begin{equation}
G_0 ({\bf p}, \tau) = e^{ - \tau p^2/2M }
\label{G0}
\end{equation}
and
\begin{equation}
g_{e, \sigma }({\bf p}, \tau >0) = -(1-n_{\bf p}) e^{ - \tau (p^2/2m - \mu ) },
\label{Ge}
\end{equation}
respectively,  where $n_{\bf p}$ is the Fermi distribution function with the chemical potential $\mu$ and electron mass $m$. The only difference with the standard electron-phonon polaron problem is that now the role of the phonon propagator is played by the $\Phi$-function. For jellium, the electron density and $r_s$ parameters are defined by standard expressions, $n=k_F^3/3\pi^2$, $r_s = (9\pi /4)^{1/3} me^2/k_F$. In this work we use units such that the chemical potential (Fermi energy)
and Fermi momentum are set to unity $\mu=\epsilon_F=1$, $k_F=1$ (i.e. $m=1/2$).

\section{Plasmon effect}

Screening of the Coulomb potential leads to collective plasmon excitations at small momenta and, correspondingly, the bosonic propagator $\Phi$ features a plasmon pole. Before proceeding with the rigorous calculation for the full dielectric function, we consider first the so-called plasmon-pole approximation for $\varepsilon$ that neglects gapless particle-hole excitations.
The idea behind PPA is to write a simple functional form that satisfies exactly two limiting cases: \\
(1) at $\omega = 0$ the static screened interaction at small momenta should have the Yukawa form, or
\begin{equation}
\varepsilon^{-1}-1 = -\frac{\kappa^2}{q^2 + \kappa^2} \, , \qquad (\omega = 0)\,,
\label{Ep_Yuk}
\end{equation}
with $\kappa^2 = 6 \pi n e^2 / \epsilon_F$; \\
(2) at $q \to 0$, and small, but finite, frequency $\omega \gg q k_F/m$ the dielectric function features a zero at the plasmon frequency
\begin{equation}
\varepsilon^{-1}-1 = \frac{\omega_p^2}{\omega^2-\omega_p^2} \, , \qquad (q = 0)\,,
\label{Ep_PPA}
\end{equation}
with $\omega^2_p = 4 \pi n e^2 / m$.  \\
These considerations lead to the following simplified PPA expression
\begin{equation}
\Phi({\bf q},\omega ) = \frac{4 \pi e^2}{q^2} \frac{\omega^2_p}{\omega^2 - \omega^2_p (1+q^2 / \kappa^2)} \,.
\label{Phi-PPA}
\end{equation}
In the imaginary time representation it reads
\begin{equation}
\Phi({\bf q},\tau) = - \frac{2 \pi e^2 \omega_p}{q^2 \sqrt{1+q^2/\kappa^2}} \; e^{ - \omega_p \sqrt{1+q^2/\kappa^2} \; \tau }.
\label{Phi-tau}
\end{equation}

\begin{figure}[tbh]
\includegraphics[width=0.9\columnwidth]{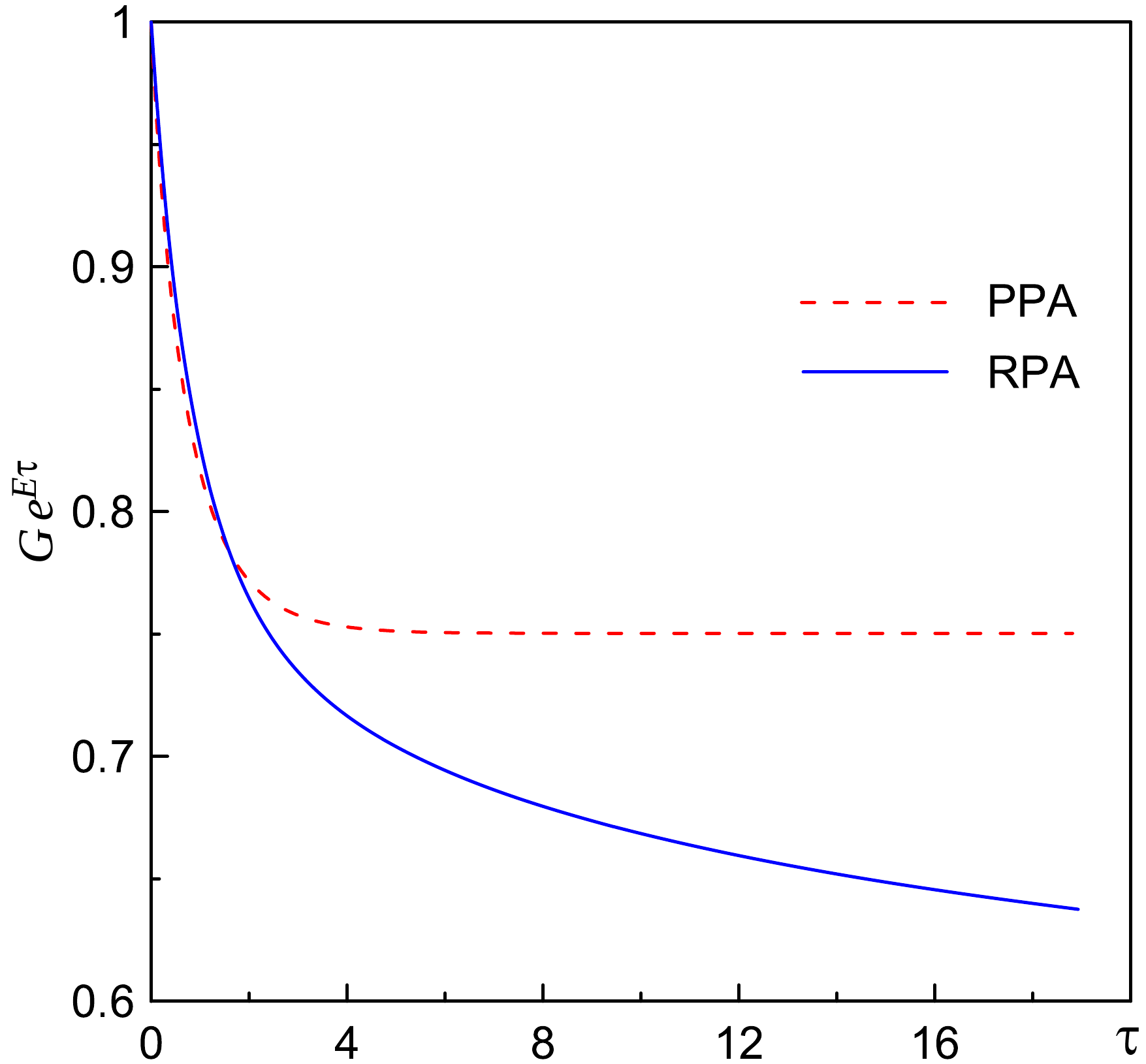}
\caption{(color online). Impurity Green's function $G$ (with the exponential dependence subtracted for clarity) in the plasmon-pole (dashed red line) and random phase (solid blue line) approximations for the dielectric function $\varepsilon$. Results are shown for $r_s=1$ and $M \to \infty$. }
\label{fig3}
\end{figure}

By substituting (\ref{Phi-tau}) into Eqs.~(\ref{Gh_Inf_1})-(\ref{Gh_Inf_2})
we find that the impurity energy
\begin{equation}
E = \int_0^{\infty}  d\tau \int \frac{d^3 q}{(2\pi)^3} \; \Phi({\bf q},\tau) \,,
\label{PolEner}
\end{equation}
and the $Z$ factor are given by
\begin{equation}
E = - e^2\kappa/2, \;\;\;\;\; Z = e^{-e^2\kappa / \pi \omega_p}.
\label{E-Z}
\end{equation}
The complete dependence on $\tau$ at $r_s=1$ is presented in Fig.~\ref{fig3} by the dashed line.

As one can see from Fig.~\ref{fig3}, in contrast with the result based on the RPA approximation to $\varepsilon$,  accounting for both gapped and gapless modes in the metal, the plasmon-pole approximation eliminates the OC. This happens because for all momenta the decay of the $\Phi$ function is exponential and controlled by the spectrum with the energy gap $\omega_p$. For the OC to take place, one needs excitations with linear density of states in the limit of vanishing excitation energy; these excitations are neglected within the PPA treatment. Nevertheless, the PPA provides a reasonable description of $G(\tau )$ at short times.

\section{Screening effect}

Even for short-range impurity potentials, $V_S$, one may wonder to what degree the OC
is modified by screening effects in the metallic system.
To this end, we compare the OC for two versions of the $\Phi$ function---with and without
screening---when the impurity potential can be treated perturbatively. Without screening,
the $\Phi$ function is based on the first diagram in Fig.~\ref{fig1}(a).
To account for screening, we consider the entire series shown in Fig.~\ref{fig2}(a) where we replace
the first and the last potentials (at $\tau_1$ and $\tau_2$) with $V_S$.
For this comparison, we take $V_S = 4 \pi e^2 / (q^2 + \kappa^2)$; the exact form of the
short-range potential $V_S$ is of little relevance here.

\begin{figure}[tbh]
\includegraphics[width=0.9\columnwidth]{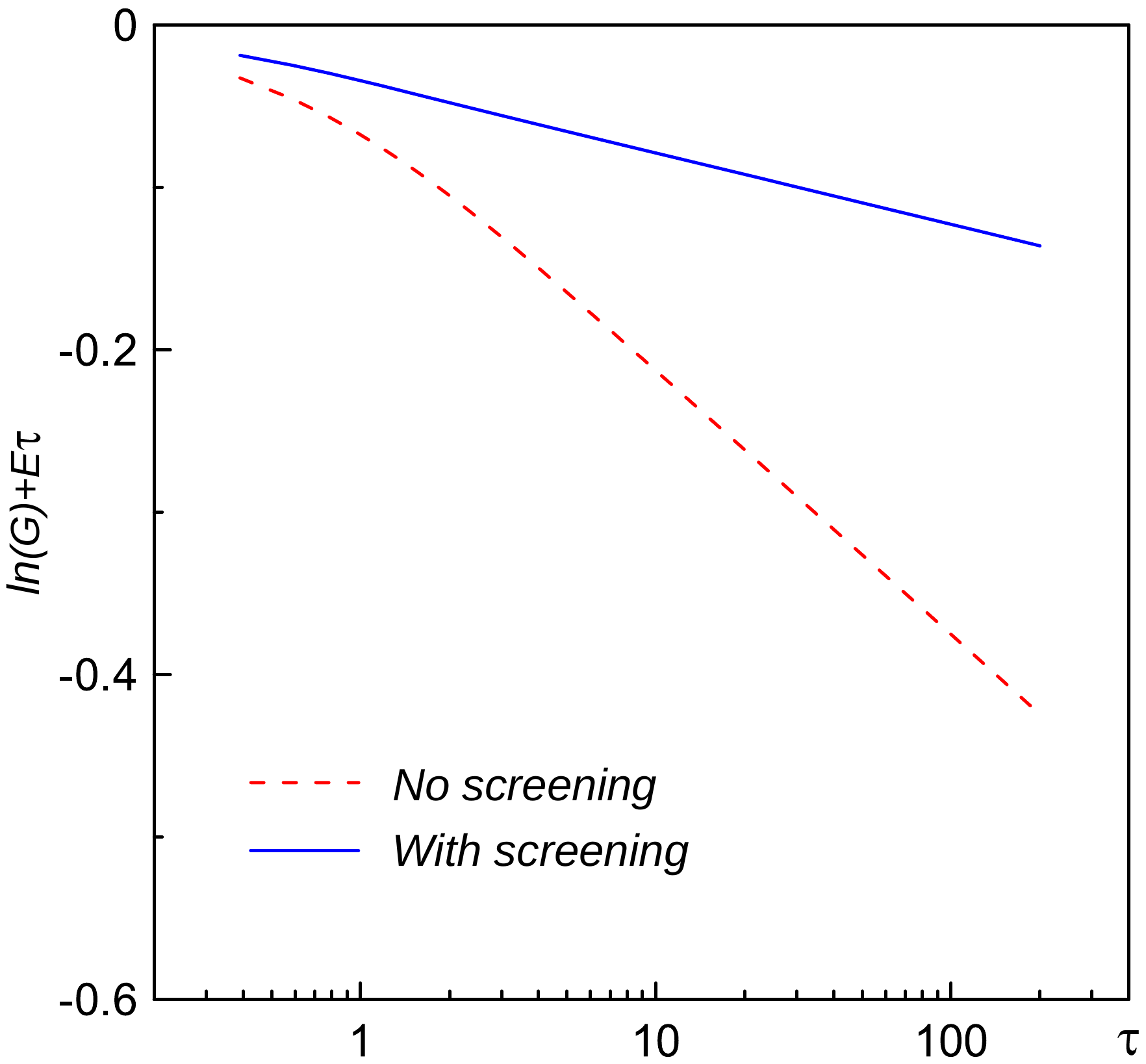}
\caption{(color online). $\ln(G)$ (with the exponential dependence subtracted for clarity from both curves) for two scenarios, with (blue solid line) and without (red dashed line) screening (see text).
Results are shown for $r_s=1$, $M \to \infty$.}
\label{fig4}
\end{figure}

The result of calculations based on Eqs.~(\ref{Gh_Inf_1})-(\ref{Gh_Inf_2}) is presented in Fig.~\ref{fig4}. To reveal the OC more clearly, we add the $E \tau$ dependence to $\ln (G)$, which is the dominant contribution to the exponent at long times, see Eq.~(\ref{PolEner}). The plot for $\ln (G) + E\tau$ without screening reproduces the standard power-law answer for OC (note the logarithmic scale for $\tau$ in Fig.~\ref{fig4}). When screening effects are accounted for, the power-law decay at long time scales has an exponent that is strongly reduced (by more than a factor of three for $r_s=1$) relative to the non-interacting Fermi-sea result. This clearly invalidates the perception that residual interactions between quasiparticles in the metallic Fermi-liquid regime can be neglected.

\section{Orthogonality catastrophe for charged impurities in metals and the finite mass effect}

Consider now the most interesting case when all potentials entering the $\Phi$ function
are of the Coulomb type, as in Fig.~\ref{fig2}(a). For finite $M$ we can no longer rely on
Eqs.~(\ref{Gh_Inf_1})-(\ref{Gh_Inf_2}) and need to employ the DiagMC technique instead.
At the formal level, the entire setup is identical to that for Frohlich polarons,
see Fig.~\ref{fig2}(b) and Refs.~\cite{Polaron1998,Polaron2000}, with the proper replacement
of the phonon propagator with the $\Phi$ function.

\begin{figure}[tbh]
\vspace{-5mm}
\includegraphics[width=0.9\columnwidth]{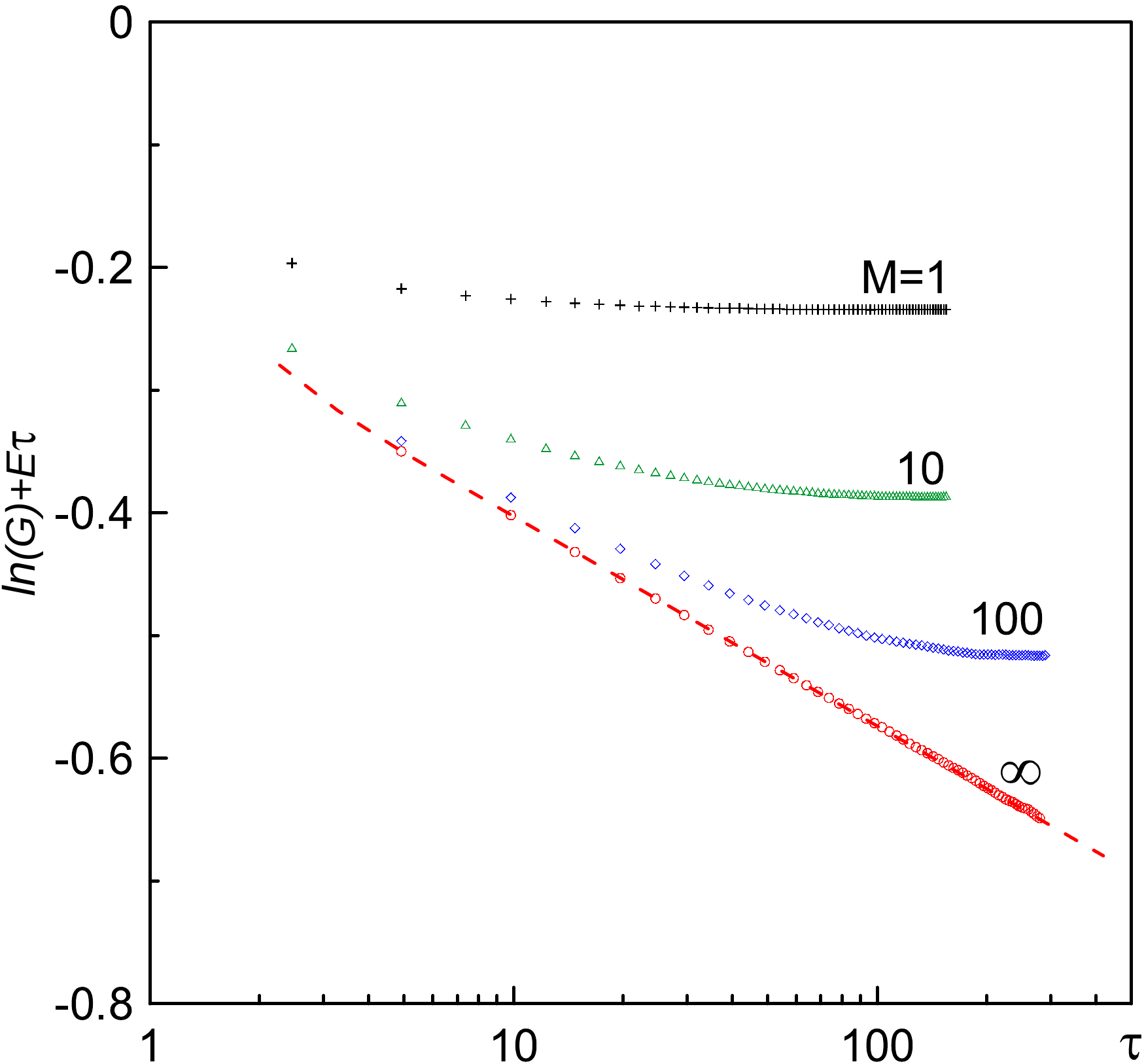}
\caption{(color online).
$\ln(G) + E\tau $ for localized (red circles) and mobile
($M = 100$---blue diamonds, $M = 10$---green triangles: $M = 1$---black crosses) impurities.
Error bars are smaller than symbol sizes. $M \to \infty$ data were benchmarked against the prediction of Eqs.~(\ref{Gh_Inf_1})-(\ref{Gh_V}) shown by the red dashed line.
The Coulomb parameter was set to $r_s=1$.}
\label{fig5}
\end{figure}

In Fig.~\ref{fig5} we show $\ln (G) + E \tau $ as a function of $\tau$ for different impurity masses. For localized impurity, as expected, we observe that $Z(\tau)$ decays to zero according to the power-law, $\propto \tau^{- \gamma}$. Screening eliminates the $q \to 0$ divergence and ensures that the integral over momentum transfer in Eq.~(\ref{Gh_Inf_2}) is finite for finite $\tau$.
However, the dielectric function retains the contribution from gapless electron-hole excitations near the Fermi surface, and these modes ultimately result in the standard power-law OC scenario for the overlap integral. By comparing Figs.~\ref{fig4} and \ref{fig5} one can see that the power-law decay exponent $\gamma$ in the case of dynamically screened Coulomb potential is significantly larger than that for the Yukawa potential in the non-interacting Fermi gas.

For finite $M$, the overlap integral is expected to saturate to a constant because the logarithmic divergence of the double integral over time in Eq.~(\ref{Gh_Inf_1}) is truncated at the inverse impurity recoil energy. As the impurity mass is getting lighter, the domain of the power-law decay in $Z(\tau)$ shrinks and ultimately reaches short time scales $ \sim \epsilon_F^{-1}$, eliminating all signatures of the Fermi-edge singularity. This behavior has been demonstrated for the case of short-range potentials in $3D$ (see, for instance, Refs.~\cite{Recoil-I,Recoil-II}). In Fig.~\ref{fig5} we show how this physics is playing out for the dynamically screened Coulomb potential.

\section{Conclusions}

We addressed the fundamental problem of the orthogonality catastrophe in Coulomb systems. For short range potentials in non-interacting Fermi gases it was solved half a century ago, but the key assumptions of the standard theory (impurity potential with finite cross-section and non-interacting fermions \cite{Mahan1967,Anderson1967,Nozier-I,Nozier-II,Nozier-III,Nozier1971}) do not apply to problems involving charged particles in metals.

We systematically investigated the OC for dynamically screened Coulomb interactions and quantified effects of gapped plasmon excitations, dynamic screening, and finite impurity mass.
While the OC retains its universal power-law decay character thanks to the gapless particle-hole excitations across the Fermi surface, the exponent $\gamma$ is sensitive to screening effects and is subject to the non-perturbative renormalization in metals. For dynamically screened Coulomb potential, we found that $\gamma$ is larger than the prediction of the phenomenological treatment based on the static Yukawa potential in the non-interacting Fermi gas. For short-range potentials, screening effects dramatically decrease the value of $\gamma$. We also found that recoil effects for finite impurity mass eliminate the OC for light particles, and convert it to small, but finite, impurity $Z$ factors for heavy particles.

The semi-analytical approach based on Eqs.~(\ref{Gh_Inf_1})-(\ref{Gh_V}) allows one to study the OC phenomenon for a variety of localized impurity problems. For mobile impurity, one has to employ the Diagrammatic Monte Carlo technique, and the most promising general algorithm for fermionic environments would be the determinant approach. However, for small values of $r_s$ the problem is reduced to the standard ``Bose-polaron" formulation where particle-hole excitations in the Fermi liquid play the role of effective bosonic medium. Future work should address finite temperature properties of such polarons.

\section{Acknowledgements}

We thank A. Tsvelik and L. Pollet for discussions. This work was supported by the Simons Collaboration on the Many Electron Problem, the National Science Foundation under the Grant No. PHY-1720465, and the MURI Program ``New Quantum Phases of Matter" from AFOSR.

\end{document}